# Large Coherence Area Thin-Film Photonic Stop-Band Lasers

Victor I. Kopp

*Department of Physics, Queens College of CUNY, Flushing, New York 11367*

*Chiral Photonics, Inc., New York, New York 10002*

Zhao-Qing Zhang

*Physics Department, Hong Kong University of Science and Technology, Clear Water Bay, Kowloon, Hong Kong*

Azriel Z. Genack

*Department of Physics, Queens College of CUNY, Flushing, New York 11367*

*Chiral Photonics, Inc., New York, New York 10002*

We demonstrate that the shift of the stop band position with increasing oblique angle in periodic structures results in a wide transverse exponential field distribution corresponding to strong angular confinement of the radiation. The beam expansion follows an effective diffusive equation depending only upon the spectral mode width. In the presence of gain, the beam cross section is limited only by the size of the gain area. As an example of an active periodic photonic medium, we calculate and measure laser emission from a dye-doped cholesteric liquid crystal film.

Photonic band gap materials hold promise as a platform for a new generation of efficient, compact photonic devices. The initial excitement generated for these materials has been sustained by the discovery of new physical effects leading to new applications [1]. In this Letter, we present an unexpected conjunction of strong expansion of the region of phase coherence within the medium and weak divergence of the output beam, which is made possible in a photonic stop band structure. Our calculations show that a diffusion-like spreading of the coherent field leads to wide area lasing at the edge of a photonic stop band or at a localized state in the middle of the stop band. These results are supported by measurements of the spatial profile of laser emission from a promising organic material – a dye-doped cholesteric liquid crystal [2, 3]. Since the maximum excitation energy is proportional to the laser area, large-area thin-film devices provide a new approach for high-power lasers. These enables lightweight optical sources for free-space communications, coherent backlighting for 3-D holographic and projection displays, and therapeutic irradiation of large areas of skin.

Periodic dielectric materials in one, two or three dimensions may possess a gap in the photon density of states in the corresponding number of dimensions [1]. In 3-D periodic structures with sufficient modulation in the dielectric function, a full photonic band gap may appear in which spontaneous emission is suppressed.

Disturbing the periodicity of an active crystal introduces long-lived localized modes into the photonic band gap at which lasing is facilitated. Though lasing has not yet been observed in 3-D periodic structures, the enhanced photon dwell times and microcavity effects in localized states have suppressed the laser threshold in lower dimensional periodic systems. Unlike lasers based on Fabry-Perot resonators, the band-edge and defect modes of 1-D structures are significantly separated from adjacent modes and have narrower linewidth than other modes. They may therefore operate as single mode lasers over a significant power range. The shift of the band gap with oblique angle at the band-edge frequency only permits the propagation of light with k-vector components near the normal. Unlike conventional spreading of light between two mirrors, which requires oblique k-vector components (Fig. 1a), photonic band structures lead to enhanced spreading of light with only near-normal k-vector components (Fig.1b).

It would appear that the best candidate for a 1-D periodic structure, for large area lasing would be a vertical cavity surface emitting laser (VCSEL) [4]. But high order transverse modes arise [5] in small-diameter VCSEL, while in large-diameter VCSELs spontaneous filamentation results from structural nonuniformities [6]. Another class of periodic one-dimensional structures, which have recently been produced as macroscopic single domain materials, is that of self-organized polymeric cholesteric liquid crystals (CLCs) [7]. Lasing at the edge of the stop band of CLCs has recently been demonstrated [2] and here we will specifically consider measurements and simulations of these materials.

The molecular director in cholesteric structures lies in a molecular plane and executes a helical rotation perpendicular to the planes. A stop band centered at a wavelength in the medium equal to the helical pitch $P$ is created for waves propagating along the helical axis. Optically pumped lasing in dye-doped CLCs occurs at the first mode at the edges of the stop band at vacuum wavelengths of $n_eP$ or $n_oP$, where $n_e$ and $n_o$ are the extraordinary and ordinary indices of refraction, respectively. The narrowest mode in CLCs is the first mode on the high frequency side of the band, which is composed of two counterpropagating components travelling normal to the layers of the structure and has an electric field everywhere aligned with the lowest index axis of the optical indicatrix. Consequently lasing generally occurs only in this mode slightly above threshold. For off-normal incident radiation, the gap shifts to higher frequency, placing oblique waves at the frequency of the lasing mode inside the stop band. As a result, transmitted or emitted radiation is confined to a narrow range of angles about the normal direction even for an incident beam with a large angular divergence as shown in Fig.1 b. The restricted angular distribution corresponds to a wide beam at the output surface of the thin film and hence to a beam inside the sample with divergence greatly exceeding that of free space diffraction divergence.

The CLC film studied was a right-handed structure with $n \sim 1.7$ and $\mathbf{D}n \sim 0.2$ and a thickness of 35 μm. The samples were doped with laser dye PM-597 with an absorption peak at 530 nm and an emission peak at 590 nm and pumped with 150 ns pulses at the second harmonic of Q-switched pulses of a Nd:YAG laser at 532 nm. Lasing is produced at the mode closest to the upper band edge. The emitted beam is perpendicular to the CLC film and produces the pattern on a screen 11.5 cm from the sample shown in the insert in Fig. 3.

The CLC medium is modeled as a set of anisotropic amplifying layers with thickness $h$ significantly smaller than the wavelength of the incident light. The direction of the molecular axis is rotated between successive layers by a small angle $2\pi P/h$. A normally incident circularly polarized 1-D Gaussian beam with the helicity of the CLC structure and frequency of the lasing mode is incident upon the sample. For simplicity of calculation, we consider an incident beam, which is a superposition of waves with an intensity distribution that is Gaussian in the plane of incidence (the *x-z*-plane) and is homogeneous in the perpendicular direction. The main results obtained for this wave are readily generalized to the case of a Gaussian beam. Transmission is calculated using a 4 × 4 transfer matrix method [8, 9, 10] for each Fourier component of the beam which are then superposed to give the transmitted wave.

The transfer-matrix is computed numerically for each plane wave component by taking the product of the transfer matrix for each layer of thickness $h$ with uniform dielectric tensor. The value of $h$ is decreased until the value of the transmittance converges. Transmission is simulated in a CLC film with $P = 370$ nm, thickness $L = 35$ μm and extraordinary and ordinary dielectric constants $\mathbf{e}_e$ and $\mathbf{e}_o$ which correspond to refractive indices $n_e = 1.8$ and $n_o = 1.6$. For the background medium, we use $n_b = 1.52$, which equals the refractive index of the glass substrates. Gain is introduced by adding a negative imaginary part to both $\mathbf{e}_e$ and $\mathbf{e}_o$. The width of the band-edge mode in the transmission spectrum drops with increasing gain and vanishes at a critical value of the gain at the laser threshold. At this point, both the transmittance and reflectance diverge.

The simulations show that the wavefunction at the output surface for a gain coefficient slightly below the critical value is proportional to $\exp[-\mathbf{a}(1-i)|x|]$ at the frequency of the band edge state for the perpendicularly propagating wave. A typical intensity distribution is shown in linear and logarithmic plots in Fig. 2. The intensity distribution is seen to decay exponentially, $I \propto \exp(-2\mathbf{a}|x|)$. Except for a small region near the beam center, the phase of the electric field increases linearly with the same coefficient $\mathbf{a}$, i.e., $\mathbf{f} = \mathbf{a}|x|$. The value of α decreases with increasing gain coefficient and vanishes at the critical gain. The value of $\mathbf{a}$, and hence the wavefunction, is found to be independent of the width of the incident beam.

The calculated intensity distribution at the output surface gives rise to an oscillatory structure in the far field. The measured and calculated far-field intensities are shown in Fig. 3 at a distance of 11.5 cm from the sample. The agreement is good outside of the central region, but calculations give a flatter peak than is measured. This is due to the use of a beam with a Gaussian intensity distribution only in one direction and to the finite gain region created by the pump beam. We now demonstrate analytically that spatial distribution found is a direct consequence of the Lorentzian shape of the resonant peak of the transmitted electric field,

$$T_b(k) \approx \frac{C_b}{B - i(k - k_r)} ; \quad \mathbf{b} = x, y, z , \qquad (1)$$

where $C_b$ and $B$ are constants, $k = 2\pi n_b / \lambda$ and $k_r = 2\pi n_b / \lambda_r$ are wavevectors in the embedding and resonant medium, respectively. The linewidth of the transmission spectrum is $\Delta\lambda = B\lambda_r^2/2\pi n_b$. When the gain coefficient $\mathbf{g}$ is slightly below its critical value $\mathbf{g}_c$, $B$ is proportional to $(\mathbf{g}_c-\mathbf{g})$. Equation (1) can be generalized to include oblique incident waves if we replace $k$ by $k \cos A$. At the band-edge frequency $k = k_r$, the transmittance is strongly peaked in the normal direction. In the limit of $B \ll k_r$, we can expand $\cos A \cong 1 - A^2/2$, and Eq. (1) gives,

$$T_b(A, k = k_r) \approx \frac{2C_b}{2B + iA^2 k_r}. \qquad (2)$$

Using Eq. (2) and superposing all transmitted plane waves we obtain the transmitted electric field:

$$E_b^T(x, z = L) \propto \exp[-(1-i)\sqrt{Bk_r}\,|x|]. \qquad (3)$$

This is precisely the wavefunction indicated in Fig. 2 with $\mathbf{a} = \sqrt{Bk_r}$. The wavefunction obtained in Eq. (3) depends only on $Bk_r$ and is independent of the spatial extent of the incident beam. Since $B$ is proportional to ($g_c$-$g$), the wavefunction becomes an unbounded plane wave as the gain approaches the critical value. When $|x| \ll 1/\sqrt{Bk_r}$, Eq. (3) becomes invalid and it cannot describe the central region of the beam.

Using Eq. (3), the width of the beam is defined as $2x_0$, where $x_0$ is the position at which the intensity drops to half of its peak value. This gives a beam width at the output surface of $W \equiv 2x_0 = \ln 2/\sqrt{Bk_r}$. By substituting $2\Delta\lambda = B\lambda_r^2/\pi n_b$ and $k_r = 2\pi n_b/\lambda_r$ into the previous relation, we find a universal relation between $W$ and the linewidth $2\Delta\lambda$ in transmission for normally incident radiation at $\lambda_r$,

$$\frac{\lambda_r}{n_b W} = \frac{\sqrt{2}\pi}{\ln 2}\sqrt{\frac{2\Delta\lambda}{\lambda_r}}. \tag{4}$$

This relation is valid as long as $B \ll k_r$. This condition is satisfied when the sample is sufficiently thick or ($g_c - g$) is sufficiently small. Since this condition does not explicitly depend on the sample characteristics, we expect that it holds also for lasing modes in other layered media, such as binary-layered (BL) media and Fabry-Perot (FP) resonators. This is demonstrated in Fig. 4, where we plot $\lambda_r/n_b W$ vs. $\Delta\lambda/\lambda_r$ on a log-log scale for three arbitrarily chosen systems: (a) CLC (◊), (b) BL (+), (c) FP (x). We also plot Eq. (4) as a solid line for comparison. In the absence of gain, we also show the data for two CLC samples with different thickness (□).

The universal relation in Eq. (4) has a surprising interpretation in terms of diffusion. Since the linewidth in the transmission spectrum is inversely proportional to the photon dwell time $\tau$, Eq. (4) can be interpreted in terms of the diffusion relation: $W^2 \approx D\tau$, since $\tau \approx \lambda_r^2/(2\pi c\Delta\lambda)$. The effective diffusion constant is $D \approx c\lambda_r/2\pi n_b^2 = c/n_b k_r$. So, the diffusion constant is proportional to the wavelength of the lasing mode in the background medium and is independent of the sample thickness and gain coefficient. Although we have assumed the wave is homogeneous in the *y* direction, this condition can be relaxed in the analytical approach. If the incident beam is a Gaussian wave in both the x and y directions, the generalization of the 1-D approach leads to an outgoing wave: $E_b^T \propto (1/\sqrt{r})\exp[-(1-i)\sqrt{Bk_r}\, r]$, where $r = \sqrt{x^2 + y^2}$ and again the condition $r \gg 1/\sqrt{Bk_r}$ should be satisfied. This is similar to the wavefunction found in Eq. (3). These results may be naturally applied to band-edge or localized modes in periodic 1-D structures since only a singe mode of radiation exists over a wide angular range centered on the normal direction to yield wide area thin-film lasers.

**Acknowledgements**

This work was supported by the National Science Foundation under Grant No. DMR9973959, and by the Army Research Office. It was performed in part under the auspices of the City University of New York CAT in Ultrafast Photonic Materials and Applications.

**Figure Captions**

Fig. 1. Coherent spreading of beam inside an amplifying medium. **a** Two flat mirrors define a cavity, in which spreading of the beam involves oblique k-vector components, which correspond to losses from the laser mode. The resulting beam size is linearly proportional to the photon dwell time. **b** Layered photonic band-gap medium, in which all oblique components of the k-vector are reflected, since their propagation is forbidden, except for k-vector component close to the normal. The beam size follows a diffusion equation and is proportional to the square root of the photon dwell time.

Fig. 2. Computer simulation of spatial distribution of laser emission at output surface of CLC. **a** Intensity and phase in linear scale; **b** Intensities of incoming and outgoing beams in semi-log scale.

Fig. 3. Spatial distribution of laser emission in far field (11.5 cm from sample).

Fig. 4. Universal relation of inverse beam width to relative linewidth for different samples (Eq. 4).

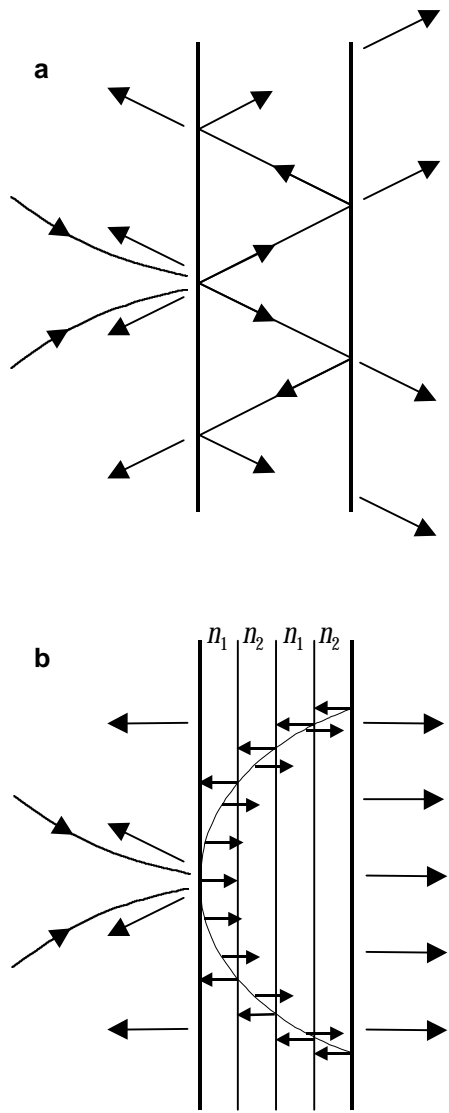

Fig. 1.

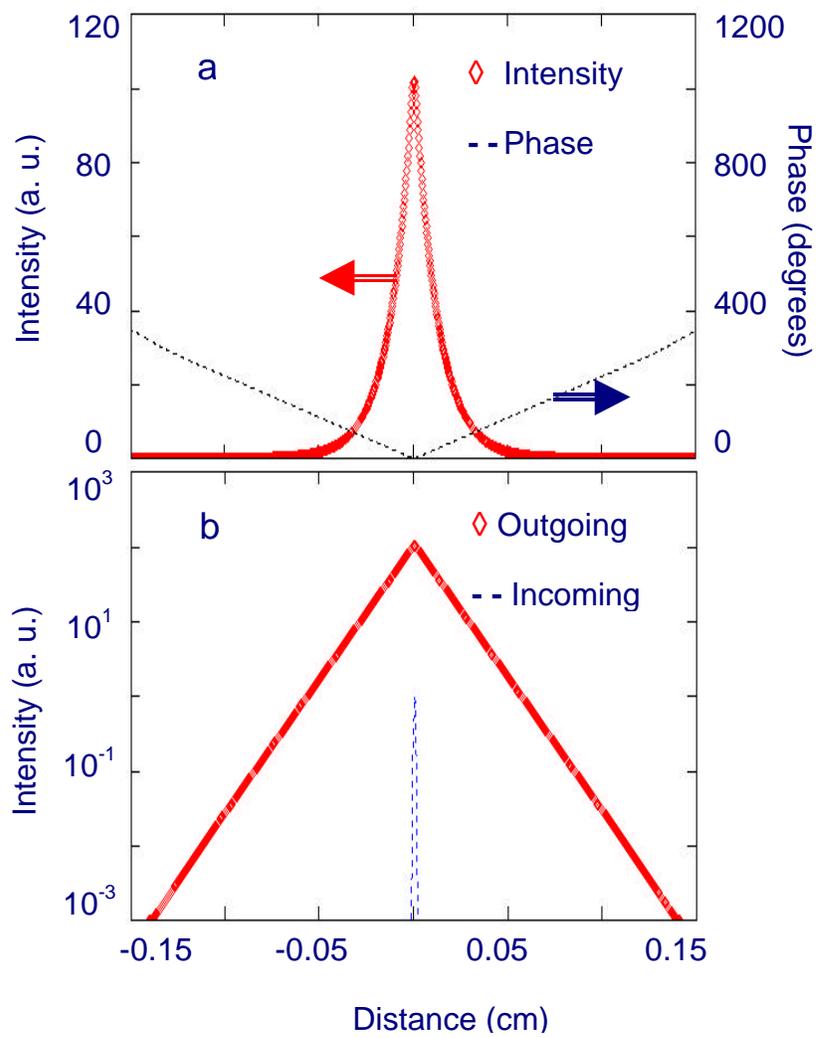

Fig. 2.

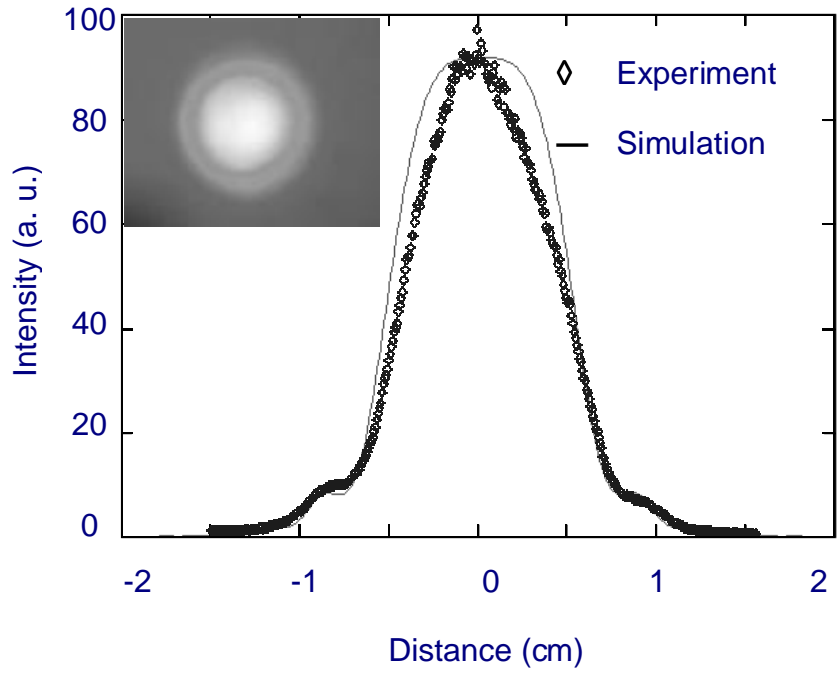

Fig. 3.

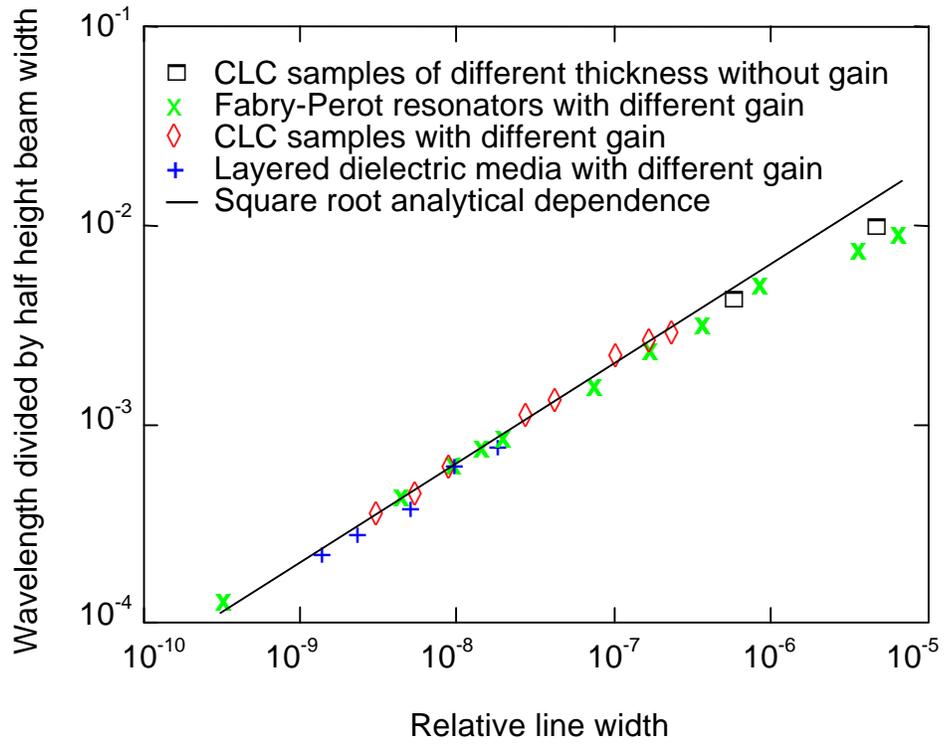

Fig. 4